\documentclass[aps,preprint,showpacs,floatfix]{revtex4}
\usepackage{graphicx,bm,amsmath,dcolumn}
\graphicspath{{figures}}
\begin{document}
\newcommand{\half}{\frac{1}{2}}
\newcommand{\ith}{^{(i)}}
\newcommand{\im}{^{(i-1)}}
\newcommand{\gae}
{\,\hbox{\lower0.5ex\hbox{$\sim$}\llap{\raise0.5ex\hbox{$>$}}}\,}
\newcommand{\lae}
{\,\hbox{\lower0.5ex\hbox{$\sim$}\llap{\raise0.5ex\hbox{$<$}}}\,}
\newcommand{\be}{\begin{equation}}
\newcommand{\ee}{\end{equation}}
\newcommand{\bea}{\begin{eqnarray}}
\newcommand{\eea}{\end{eqnarray}}

\title{Critical frontier of the Potts and percolation models on triangular-type and kagome-type
lattices I: Closed-form expressions}
 
\author{F. Y. Wu} 
\affiliation{Department of Physics, Northeastern University, Boston, Massachusetts 02115, USA}
 

\begin{abstract}
We consider the Potts model and the related  bond, site, and mixed site-bond percolation
problems on
triangular-type and kagome-type lattices, and derive closed-form expressions for the critical frontier.
For triangular-type lattices the critical frontier is known,
usually derived from a duality consideration in conjunction with the assumption of a unique transition.
 Our analysis, however, is rigorous and based on an established  result without the need 
of a uniqueness assumption, thus firmly establishing all derived results. 
For  kagome-type lattices the exact critical frontier is not known.
We derive a closed-form expression for the Potts
critical frontier by making use of a homogeneity assumption. The closed-form expression
is new, and we apply it to a host of problems including  site, bond, and 
mixed site-bond percolation on various lattices.
   It yields exact thresholds for site percolation on 
 kagome, martini, and other lattices, 
and is highly accurate numerically in other applications when compared to numerical determination.
   \end{abstract}
\maketitle 

\section{Introduction}
\label{intro}
An outstanding problem in lattice statistics is the determination of the critical frontier, or
 the loci of critical point, of lattice models. Of special interest is the 
$q$-state Potts model \cite{potts52} and its  associated lattice models \cite{wureview}.
 For $q=2$ it is the Ising
model, and for $q=1$ the Potts model generates the percolation problem \cite{wu1978} including 
the bond \cite{kas},  site \cite{kw}, ane mixed site-bond  
percolation. However, except for the simple square, triangular and honeycomb lattices \cite{wu79} 
and some special lattices essentially of a triangular-type \cite{wupotts06}, 
the determination of the Potts critical
frontier in general has proven to be elusive.  

\begin{figure}[htbp]
\includegraphics[scale=0.4]{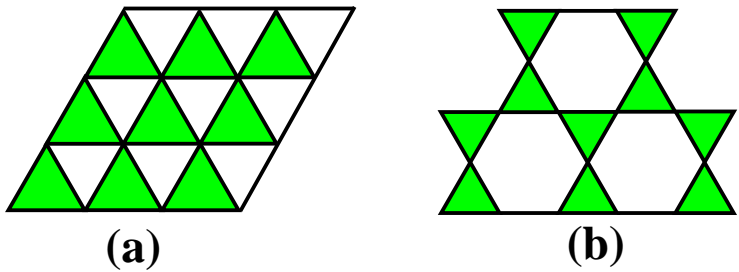}
\caption{(a) Triangular-type lattice. (b) Kagome-type lattice. Shaded triangles possess
 Boltzmann weights (\ref{ABCdef}).}
\label{threesite}
\end{figure}

 In this paper we consider the Potts model on
two general classes of lattices,
 the  triangular- and kagome-type lattices  shown in Fig. \ref{threesite}.
  Shaded triangles in Fig. \ref{threesite}
denote   interactions involving 3 Potts spins $\tau_1, \tau_2, \tau_3 = 1,2,...,q $ with the
Boltzmann weights
\begin{eqnarray}
W_\bigtriangleup (1,2,3) &=& A + B(\delta_{12}+\delta_{23}+\delta_{31})+C\delta_{123}\, , \nonumber \\
W_\bigtriangledown (1,2,3) &=& A' + B'(\delta_{12}+\delta_{23}+\delta_{31})+C'\delta_{123}\, , \label{ABCdef}
\eea
where $\delta_{ij} = \delta_{\rm Kr}(\tau_i, \tau_j), \delta_{123}=\delta_{12}\,\delta_{23}\,\delta_{31}$,
and  $A,B,C, A', B', C'$ are constants. In (\ref{ABCdef}), we have
assumed interactions isotropic in the 3 directions of a triangle. The extension of our analysis to anisotropic 
interactions is straightforward and will not be given.
 Special cases of shaded triangles are the ``stack-of-triangle", or subnet, lattices  
 shown in Figs. \ref{subnet1} and \ref{subnet2} that have been of recent interest
\cite{yao08,loh08,ziffgu09,HAZiff09}.
   We refer
to these stack-of-triangle lattices as  subnet lattices. The $1\times 1$ subnet 
lattices are   the  triangular and kagome lattices themselves.
  We shall call a kagome-type lattice with \ $m\times m$ \ down-pointing  and \ $n\times n$
 \ up-pointing subnets an \ $(m\times m):(n\times n)$ \ subnet lattice,
or simply an \ $(m\times m):(n\times n)$ \ lattice.  Examples of these kagome-type subnet lattices
are shown in Fig. \ref{subnet2}.
 
\begin{figure}[htbp]
\includegraphics[scale=0.4]{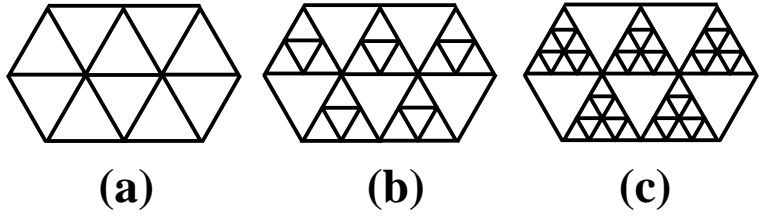}
\caption{Triangular subnet lattices. (a) $1\times 1$ lattice
(triangular). (b) $2\times 2$ lattice. (c) $3\times 3$ lattice.}
\label{subnet1}
\end{figure}

\begin{figure}[htbp]
\includegraphics[scale=0.4]{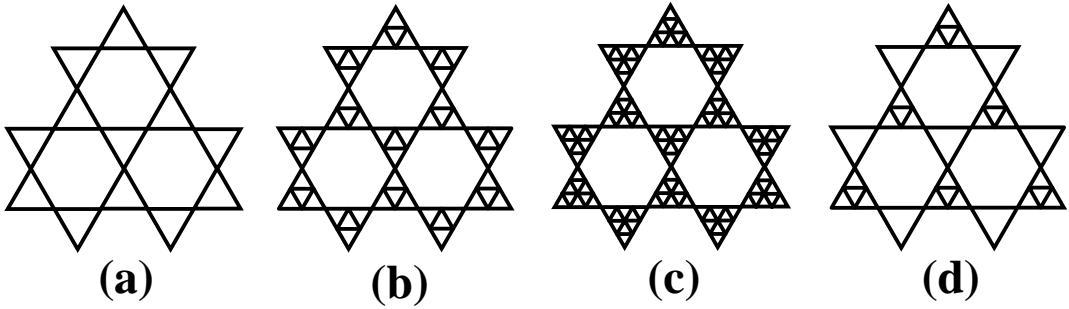}
\caption{Kagome subnet lattices. (a) $(1\times 1):(1\times 1)$ lattice
(kagome). (b) $(2\times 2):(2\times 2)$ \ lattice. (c) $(3\times 3):(3\times 3)$ \  lattice.
(d) $(1\times 1):(2\times 2)$ \ lattice.}
\label{subnet2}
\end{figure}

Partition functions for the two lattices in Fig. \ref{threesite} are
\bea
Z_{\rm tri} (q; A,B,C) &=& \sum_{\tau_i=1}^q \prod_\bigtriangleup W_\bigtriangleup (i,j,k)\, , \label{triboltzmann} \\
Z_{\rm kag} (q; A,B,C; A', B', C') &=& \sum_{\tau_i=1}^q \big[\prod_\bigtriangleup W_\bigtriangleup (i,j,k)
\big] \cdot \big[\prod_\bigtriangledown W_\bigtriangledown (i', j', k') \big], \label{kagomeboltzmann}
\eea
where the products are taken over the respective up- and down-pointing shaded triangles.  
The critical frontier of the triangular-type lattice of Fig. \ref{threesite}(a) has been 
 known from earlier works \cite{bta78,wulin80,wuzia81}, but the critical frontier for the kagome-type
lattice  of Fig. \ref{threesite}(b) is open.

For  $q=2$,    we can replace
Potts spins $\tau$ by Ising spins $\sigma = \pm 1$ and  shaded triangles by triangles
($1\times  1$ subnet) with an Ising interaction $K_I$ $ (=K/2) $. To determine $K_I$, we write
\bea
\delta_{ij} &=& \frac 1 2 (1+\sigma_i\sigma_j), \nonumber \\
 W_\bigtriangleup(1,2,3)&=& {\rm exp} [2K_I(\delta_{12}+\delta_{23}+\delta_{31})]. \label{Isingint1}
\eea
 Equating (\ref{Isingint1})  with $W_\bigtriangleup$
in  (\ref{ABCdef}), one obtains after a little algebra  
\be
e^{4K_I} = (A+3B+C)/(A+B).  \label{Isingabc}
\ee
It follows that the partition functions (\ref{triboltzmann}) and (\ref{kagomeboltzmann})
are completely equivalent to those of the triangular and kagome Ising model.
 
For $q \geq 3$, the shaded triangles can be replaced by any triangular network having 2 independent 
parameters. An example is the mapping 
shown in Fig. \ref{transformation} in Sec. \ref{IIIG}.

Parameters $A,B,C$ for a given Potts subnet can be readily worked out. For the $1\times 1$ triangle,
for example, one has
\be
W_\bigtriangleup(1,2,3) = {\rm exp} [K(\delta_{12}+\delta_{23}+\delta_{31})]
\ee
from which one obtains
\be
A=1, \quad B= v, \quad C=3v^2+v^3, \quad \quad ({\rm triangle}), \label{1times1}
\ee
where $v=e^K-1$. For
the $2\times 2$ subnet, one obtains in a similar fashion
  \begin{eqnarray}
\label{net22}
A&=& 3v^5 +21v^4 +(50+4q) v^{3}+33qv^{2}+9q^{2}v+q^{3} , \nonumber\\
B&=& v^7 + 7v^6 +22v^5 +(30+2q)v^{4}+10qv^{3}+q^{2}v^{2} , \nonumber\\
C&=& v^9 +9v^8 +33v^7 +63v^{6} +(54+3q)v^{5}+9qv^4, \quad \quad (2\times 2 {\rm \>\>subnet)}.\label{2by2}
\end{eqnarray}
Expressions of $A,B,C$ for  $3\times 3$ and $4\times 4$ subnets are derived and
 given in a subsequent paper \cite{II},   hereafter referred to as II.

 The structure of this paper is as follows:
In Sec. \ref{section2} we consider the triangular  subnet lattices and 
apply the rigorously known critical frontier to  various models including mixed site-bond percolation.   
 In Sec. III we consider the 
  kagome-type lattice, and derive a closed-form expression for its critical frontier
on the basis of a homogeneity assumption. 
 We show
that this critical frontier  is exact for site percolation on the kagome, martini, and other lattices,
and  is highly accurate in  other applications. 
   The accuracy   of the critical frontier 
 will be closely examined in paper II.
 
\section{Triangular-type lattices}
\label{section2}
In this section we consider triangular-type lattices of Fig.
\ref{threesite}(a).

The Potts model  on  the triangular-type lattice  was first studied by Baxter, Temperley and Ashley \cite{bta78}
in the context of a Potts model with 2- and 3-site interactions. Using a 
Bethe-ansatz result
on a 20-vertex model on the triangular lattice due to Kelland \cite{kelland1,kelland2},
they showed that the partition function (\ref{triboltzmann}) 
 is self-dual, and derived its self-dual point which, in the language of 
the interaction (\ref{ABCdef}),  reads 
\be
qA=C. \label{tri}
\ee
This self-dual trajectory was later re-derived  graphically  by Wu and Lin \cite{wulin80}.
However, as is common in  duality arguments, an additional assumption of a unique transition is needed
to ascertain that (\ref{tri}) is indeed the actual critical frontier. 

However, Wu and Zia \cite{wuzia81} established subsequently in a rigorous analysis
that (\ref{tri}) is  indeed the
 critical frontier  in the `ferromagnetic' regime
\be
2B+C> 0, \quad
3B+C > 0 . \label{condition}
\ee
It can be verified that the condition (\ref{condition}) holds for 
(\ref{1times1}) and (\ref{2by2}),
so the critical frontier $qA=C$ is exact. 
  Applications of (\ref{tri}) to
the martini and other  lattices have been reported in \cite{wupotts06}. 
 The duality relation of the triangular Potts model with 2- and 3-spin interactions 
has also be studied by Chayes and Lei \cite{chayeslei06} with several rigorous theorems on the phase 
transition proven.
 
\subsection{Ising model}
In Sec. I, 
we have established  that for $q=2$ any triangular-type lattice  is reducible
to a triangular Ising lattice with interaction $K_I$ given by (\ref{Isingabc}).  
Indeed, using (\ref{Isingabc}),  the known critical point $e^{4K_I} = 3$ of the
triangular Ising model reduces to  the critical frontier $2A=C$ as expected.

For the Ising model on \ 
    $1\times 1$ \ and \ $2\times 2$ \ subnet lattices with interaction
$K_{I}$, we set $q=2$, $v=e^{2K_{I}}-1$  in (\ref{1times1}) and (\ref{2by2}),
and obtain from $2A=C$ the critical point
\bea
x_c &=& \sqrt 3, \quad 1\times 1{\rm \>\>subnet\>(triangular\> lattice)} \nonumber \\
         &=&  \sqrt 5, \quad 2\times 2{\rm \>\>subnet},
\eea
where $x=e^{2K_{I}}$.
Using expressions of $A$ and $C$ given in II for \ $3\times 3$ \ 
and \ $4\times 4$ \ subnets, we obtain similarly
\bea
x^8-5x^6-x^4-19x^2-8=0,\quad x_c&=&\frac 1 2 \sqrt{ 5 + \sqrt{33} +  \sqrt { (50 + 18\sqrt {33})}} \nonumber \\
&=& 2.404\ 689\ 372,  \  3\times 3{\rm \>subnet}\nonumber \\
x^{12}-5x^{10}-x^8-22x^6-53x^4-125x^2-51=0,\quad x_c&=&2.467\ 648\ 033, \ 4\times 4{\rm \>subnet}.
\eea
  
\subsection{Bond percolation}
It is well-known that bond percolation 
is realized in  the $q=1$ limit of the Potts model  under the mapping  $v=p/(1-p)$,
where $p$ is  the bond occupation probability \cite{kas,wu1978}.
Therefore the percolation threshold is given  simply by $C=A$.
Thus using (\ref{1times1}) for $A$ and $C$ for the triangular lattice, one obtains 
the well-known \cite{sykesessam,essam72,essam79} critical frontier for bond percolation   
 \be
1-3p+p^3=0, \quad {\rm or} \quad p_c=2 \sin (\pi /{18}) = 0.347\ 296\ 355. \quad {\rm (triangular\>lattice)}.
\label{bond1}
\ee

For the $2\times 2$ subnet lattice we use (\ref{2by2}) and obtain
\bea
&&1-3p^2-9p^3+3p^4+45p^5-72p^6+45p^7-12p^8+p^9=0, \nonumber \\
&& \hskip 2cm {\rm or} \quad  p_c=0.471\ 628\ 788 \quad (2\times 2 {\rm \>subnet\>\>lattice}). \label{bond2}
\eea
In a similar fashion using expressions of $A$ and $C$ given in II, we obtain
\bea
&&1 - 3p^3 - 18 p^4 - 39 p^5 + 77 p^6 + 309 
      p^7 - 198 p^8 - 1406 
      p^9 + 315 p^{10} + 9303 p^{11} - 23083 p^{12}\nonumber \\
    && \hskip 1cm  + 28707p^{13} - 22047 
        p^{14} + 10959 p^{15} - 3462 p^{16} + 636 p^{17} - 52 p^{18} = 0, \nonumber \\
   && \hskip 3cm p_c = 0.509\ 077\ 792 \quad \quad(3\times 3 {\rm \>\>subnet\>\>lattice}), \label{bond3times3}\\
&&1 - 3 p^4 - 30 p^5 - 114 
      p^6 - 63 p^7 + 636 p^8 + 1940 p^9 + 741 p^{10} - 14283p^{11} - 
      26541 p^{12} \nonumber \\ 
    &&   + 78759 p^{13} + 189279 p^{14}  
     -370589 p^{15} - 1229877 p^{16} + 2829339 p^{17} + 6938691 p^{18} - 
41655363p^{19} \nonumber \\
   &&+ 96750306 p^{20} - 143421123 p^{21} + 152405700 p^{22} - 121438416 
      p^{23} + 73822093 p^{24} - 34270647 p^{25} \nonumber \\
   &&   +11994555 p^{26}
    -   3073478p^{27} +545409p^{28}-60012p^{29}+3089p^{30} = 0, \nonumber \\
     &&\hskip 3cm p_c=  0.524\ 364\ 822  \quad  \quad (4\times 4 {\rm \>\>subnet\>\>lattice}).\label{bond4times4}
 \eea
These findings agree with  those of  Haji-Akbari and Ziff \cite{HAZiff09}
deduced from  a duality consideration. As aforementioned, our derivation now ascertains 
that these thresholds are the  exact transition points.
 
\subsection{Potts model}
The exact critical threshold for the Potts model on triangular-type lattices is 
(\ref{tri}), or $qA=C$. Using expressions of $A$ and $C$ given in (\ref{1times1}), one
obtains  the known critical frontier  \cite{kimjoseph,wu79}
 \be
3v^2+v^3=q, \quad \quad ({\rm Potts\>\>model\>\>on\>\>triangular\>\>lattice}). \label{tricri}
\ee
For the $2\times 2$ subnet lattice one uses (\ref{2by2}) and obtains the critical frontier 
\be
v^9+9v^8+33v^7+63v^6+54v^5-12qv^4-(50q+4q^2)v^3-33q^2v^2-9q^3v-q^4=0.
\label{triq}
\ee
Solutions of (\ref{tricri}) and (\ref{triq}) and those of
 the \ $3\times 3$ \ and \ $4\times 4$ \ subnet lattices are tabulated in 
Table \ref{tripotts} for $q=1, 2, 3, 4, 10$. 
 Note that the $q=1$ solutions are related to the bond percolation thresholds (\ref{bond1})-(\ref{bond4times4}) by
$e^{K_c} = 1/(1-p_c)$.

\begin{table}[htbp]
\caption{Exact Potts threshold $e^{K_c}$ for triangular-type subnet lattices.}
\begin{tabular}{c|c|c|c|c|c}
    \hline         &  $q=1$      & $q=2$ (Ising)      & $q=3$          & $q=4$   & $q=10$   \\
    \hline
     Triangular lattice \  &\ 1.532\ 088\ 885& $\sqrt 3$    &\ 1.879\ 385\ 241    &   2  &\ 2.492\ 033\ 301 \\
     $2\times 2$ \     &\  1.892\ 608\ 790  & $\sqrt 5$ & \ 2.493\ 123\ 120 &\  2.706\ 275\ 430 &\ 3.602\ 637\ 947 \\
    $3\times 3$  \ &\   2.036\ 982\ 609 &\  2.404\ 689\ 372 &\  2.674\ 398\ 828  &\ 2.895\ 419\ 068 &\ 3.808\ 005\ 450\\ 
     $4\times 4$  \   &\  2.102\ 451\ 724 &\  2.467\ 648\ 033 & \ 2.731\ 876\ 784&\  2.946\ 645\ 097&\ 3.820\ 754\ 228 \\
    \hline
\end{tabular}
\label{tripotts}
\end{table}

 \subsection{Site percolation}
\label{triangularsiteper}
Kunz and the present author \cite{kw} have shown  that
  site percolation can 
be formulated as  a $q=1$ limit of a Potts model
 with multi-site interactions. The Kunz-Wu scheme is to consider  a {\it reference} lattice
with multi-spin interactions, and regard
  faces  of multi-spin interactions as sites of a new lattice on which
the site percolation is defined. 
 The critical frontier of the Potts model on the reference lattice
then produces  the  site percolation threshold for the new lattice.
This scheme of formulation can be extended to mixed site-bond percolation. 

{\it 1. Site percolation on the triangular lattice}:

 Consider as a reference lattice the
triangular lattice  with pure 3-site interactions $M$ 
marked by dots  shown
in the left panel of Fig. \ref{trisite}.  The dots 
form a triangular lattice shown in the right. The Kunz-Wu scheme now
solves the  site percolation on the triangular lattice. 
We have
 \be
W_\bigtriangleup(1,2,3) = e^{M\delta_{123}} = 1+m\,\delta_{123}\, ,
\ee 
or $A=1, \, B=0,\, C=m=e^M-1$.  
   Writing  $m=s/(1-s)$, where $s=1-e^{-M}$ is the site occupation probability,
the exact critical frontier  $A=C$ now yields immediately the well-known 
site percolation threshold \cite{sykesessam,essam72,essam79,sudingziff99} for the triangular lattice,
\be
s_c=1/2\, . \label{trisiteperc}
\ee

\begin{figure}[htpb] 
\includegraphics[scale=0.3]{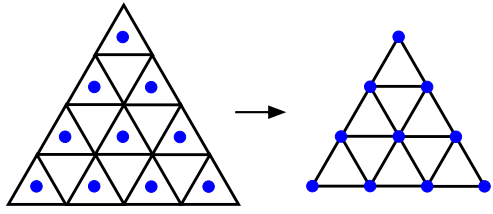}
 \caption{Site percolation on the triangular lattice.}
\label{trisite}
\end{figure}

{\it 2. Site percolation on kagome lattice}:

Consider the reference $2\times 2$ subnet lattice  with pure 3-site interactions $M$ denoted by dots shown in
 the left panel of Fig. \ref{kagomesiteper}.  The Kunz-Wu scheme then maps the reference Potts model
 into  site percolation on
 the kagome lattice as indicated in the right.  

Now for a \ $2\times 2$ \ subnet containing  3 dots as in Fig. \ref{kagomesiteper}, we  have
\be
A=q^3+3\,qm, \quad B=m^2, \quad C=m^3, \label{3dots}
\ee
where $m=e^M-1$. 
Writing $m=s/(1-s)$ and setting $q=1$,  the rigorous critical frontier 
 $A=C$  yields the critical condition $1-3s^2+s^3=0$, leading to the 
known exact result  \cite{sykesessam,essam72}
\be
s_c^{kag}= 1-2\sin(\pi/18) = 0.652\ 703\ 644. \label{kagsitecriticalpoint}
\ee

\begin{figure}[htpb]
\includegraphics[scale=0.3]{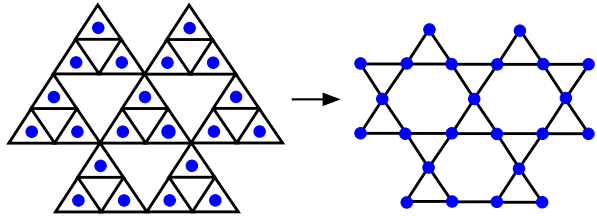}
\caption{Site percolation on the kagome lattice.}
\label{kagomesiteper}
\end{figure}

{\it 3. Site percolation on   \ $(1\times 1):(n\times n)$ \  lattices}:

The above scheme of mapping can be extended to 
 site percolation on \ $(1\times 1):(n\times n)$ \ lattices for general $n$.
The  example of Fig. \ref{kagomesiteper} 
is $n=1$, and the $n=2$ lattice  is shown in Fig. \ref{kagomesite12}.  The
reference lattice (not shown) for $n=2$ consists of  \ $3\times 3$ \ subnets with 
\bea
A&=& q^7+6q^5m+15q^3m^2+(14q+3q^2)m^3+3m^4 ,\nonumber \\
B&=& q^2 m^3 +2(q+1) m^4 +m^5 ,\nonumber \\
C&=& 3m^5+m^6,
\eea
where $m=e^M-1$, $M$ is the 3-site interaction. 
After setting $q=1$ and $m=s/(1-s)$, the critical frontier 
$qA=C$ becomes
\be
(1-2s^2)(1 + 2s^2 - 3s^3 + s^4) =0,
\ee
yielding the exact threshold 
\be
s_c = 1/\sqrt 2 ,    \quad\quad (1\times 1):(2\times 2) {\rm \>\>kagome\>\>site\>\>percolation}.
\ee
The exact critical threshold for higher \ $(1\times 1):(n\times n)$
lattices  can be similarly worked out.

\begin{figure}[htpb] 
\includegraphics[scale=0.2]{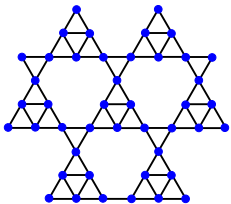}
 \caption{Site percolation on the \ $(1\times 1):(2\times 2)$ \  lattice.}
\label{kagomesite12}
\end{figure}

\section{Kagome-type lattices}
We consider in this section the case of the kagome-type lattices of Fig. \ref{threesite}(b).

 The critical frontier of the Potts model on kagome-type lattices  
has proven to be highly
 elusive. On the basis of a homogeneity assumption, however,
 the present author \cite{wu79}  has advanced a conjecture on the critical point for the kagome lattice. 
The conjecture
has since been closely examined \cite{wuhu,ziffsuding97,scullardziff,fengdengbloete08}
and found to be extremely accurate.
Here we extend the homogeneity assumption  to general
     kagome-type Potts lattices. 
For continuity of reading,  we  first state our result in Sec. \ref{conjecturesection}, 
and present the derivation in Sec. \ref{IIIG}.

\subsection{A closed-form critical frontier and homogeneity assumption} 
\label{conjecturesection}  
For  Potts model on kagome-type lattices  described by the  partition function
(\ref{kagomeboltzmann}), the
  critical frontier under a homogeneity assumption  is   
 \be
(q^2A+3q\,B+C)(q^2A'+3q\,B'+C')-3(qB+C)(qB'+C') -(q-2)CC'=0 .\label{conjecture}
\ee
\noindent
{\it Remarks:} 

1. Despite  its appearance, the critical frontier (\ref{conjecture}) actually contains only 3 
independent parameters (Cf. ({\ref{dualinteraction}) below).

2. The expression (\ref{conjecture}) is  exact for $q=2$.

\subsection{Ising model}

We first show that (\ref{conjecture}) is exact for $q=2$. We have already established
that the partition function
(\ref{kagomeboltzmann}) is precisely that of  the kagome Ising model.  For completeness we now
verify that the critical frontier (\ref{conjecture}) also gives the known kagome critical point.

For symmetric weights $A=A', B=B', C=C'$, the kagome Ising model has a uniform interaction $K_I$ 
and the  critical point  is known to be at $e^{4K_I}= 3 + 2\sqrt 3$ \cite{syozi}.
 It is readily verified that, by using (\ref{Isingabc}) for $K_I$,
 this critical point gives rise to precisely  the  $q=2$ 
 critical frontier (\ref{conjecture}), namely,
\be
4A+6B+C = \sqrt 3 \,(2B+C). \label{kagIsing}
\ee
The proof can  be extended to the kagome-type model with   asymmetric weights.

Critical thresholds of kagome-type 
Ising subnet lattices computed from (\ref{kagIsing}) are tabulated in Table \ref{asymmkagomepotts}.
For the kagome and \ $(2\times 2):(2\times 2)$ \ Ising lattices, for example, 
 we use (\ref{1times1}) and (\ref{2by2}) for $A,B,C$ with $q=1, \, v=x-1,\, x=e^{2K_I}$.
This gives 
\bea 
x^4 -6x^2-3=0, \quad x_c &=&  \sqrt {3+2\sqrt 3} \nonumber\\
                  &=&  2.542\ 459\ 756, \quad \quad \quad {\rm (kagome\> lattice)},   \nonumber \\
x^8-8x^6-6x^4-32x^2-83=0,\quad  x_c &=& \sqrt {2+\sqrt 3 +\sqrt{12+10\sqrt 3 }} \nonumber \\
                             &=&  3.024\ 382\ 957, \quad (2\times 2):(2\times 2) {\rm \>\>lattice}. \label{kagomeising}
\eea
\subsection{Bond percolation}
\begin{table}[htbp]
\caption{Bond percolation threshold $p_c$ for \ $(m\times m):(n\times n)$ \  lattices  for $m,n \leq 4$.}
\begin{tabular}{c|c|c}
    \hline    Lattice      &  This work   &\ \  Numerical determination  \\
    \hline
      Kagome    & \  0.524\ 429\ 717 \    & \ \ 0.524\ 404 \ 99(2) \cite{fengdengbloete08} \\
   ($1\times 1): (2\times 2)$  \   & \ \  0.570\ 882\ 620 \ & \ \ 0.570\ 866\ 51(33) \cite{ziffgu0910} \\
     $(1\times 1):( 3\times 3)$  \   & \ \   0.599\ 798\ 340\   &       \\
     $(1\times 1):( 4\times 4)$ \    & \ \  0.592\ 017\ 120 &       \\
     \hline

     $(2\times 2):( 2\times 2)$ \    & \ \  0.600\ 870\ 248   &  \ \ 0.600\ 862\ 4(10)   \cite{ziffgu09}   \\
     $(2\times 2):( 3\times 3)$ \    & \ \   0.610\ 916\ 740 \  &       \\
     $(2\times 2):( 4\times 4)$ \    & \ \  0.614\ 703\ 624 \   &       \\
     \hline

     $(3\times 3):( 3\times 3)$ \   & \  \ 0.619\ 333\ 485 \   & \ \ 0.619\ 329\ 6(10) \cite{ziffgu09}      \\
     $(3\times 3):( 4\times 4)$ \    & \ \  0.622 473 191  \   &       \\
      \hline

      $(4\times 4):( 4\times 4)$ \    & \ \  0.625\ 364\ 661 \   &  \ \ 0.625\ 365 (3) \cite{ziffgu09}     \\
     \hline
\end{tabular}
\label{pctab}
\end{table}

For bond percolation threshold on kagome-type subnet lattices, we again use (\ref{conjecture}) 
   with the substitution of 
$q=1$ and $v=p/(1-p)$, where $p$ is the bond
occupation probability. Using (\ref{1times1}) and (\ref{2by2}), we obtain  
  \be
1 - 3 p^2 - 6 p^3 + 12 p^4 - 6 p^5 + p^6 = 0,\quad p_c= 0.524\ 429\ 717{\rm \>\>(kagome\> \>lattice)}, \label{kagome}
\ee
\bea
1- 3p^3 - 12p^4 - 12p^5 + 63p^6 + 60 p^7 &-& 
    330p^8 + 423 p^9 -  264 p^{10}  + 84p^{11} -11p^{12} = 0,\nonumber \\
   p_c&=& 0.570\ 882\ 620, \quad (1\times 1):(2\times 2){\rm \>\>lattice},
    \label{12kagomebond} 
\eea
\bea
1 - 3p^4  - 18p^5 - 39 p^6 + 30 p^7 &+& 273 p^8+ 264 p^9 - 1785 p^{10} - 126 p^{11} + 8232p^{12} 
-  16236 p^{13} \nonumber \\
 &+& 16359 p^{14} - 9948 p^{15}+  3708 p^{16}
  - 786 p^{17} + 73 p^{18} = 0,  \nonumber \\ 
p_c&=& 0.600\ 870\ 248, \quad 
               (2\times 2):(2\times 2) {\rm \>\>lattice} . \label{kagome1}
\eea
Bond percolation thresholds computed from (\ref{conjecture}) 
for \ $(m\times m):(n\times n)$ \  lattices
are tabulated in
   Table \ref{pctab}.   We also include in Table \ref{pctab}  numerical determinations of $p_c$ for the  
\ $(1\times 1):(2\times 2)$ \cite{ziffgu0910} and \ $(n\times n)\times
(n\times n)$, $n=2,3,4$ \cite{ziffgu09}
lattices   by Ziff and Gu  using simulations, and of the kagome lattice 
by Feng, Deng and Bl\"ote \cite{fengdengbloete08} from a transfer matrix analysis.
  The comparison shows that
(\ref{conjecture})  is  accurate to within one part in $10^5$.

\subsection{Potts model}
\begin{table}[htbp]
\caption{Potts  threshold $e^{K_c}$  for kagome-type subnet lattices.}
\begin{tabular}{c|c|c|c|c|c }
    \hline    Lattice  & $q=1$    &  $q=2$ (Ising) & $q=3$ & $q=4$ & $q=10$ \\
    \hline
      kagome lattice   & \ \ 2.102\ 738\ 619\ & \ \  2.542\ 459\ 757\  & \ \ \ 2.876 269 226 & \ \ 3.155\ 842\ 236 &\ \ 4.355\ 385\ 241\\
$(1\times 1):( 2\times 2)$&\ \ 2.330\ 364\ 713\ &\ \ 2.821\ 281\ 889\ &  \ \ 3.186\ 678\ 923 & \ \ 3.489\ 096\ 458 &\ \ 4.761\ 529\ 399\\
$(1\times 1):( 3\times 3)$& \ \ 2.498\ 740\ 260\ &\ \ 2.903\ 273\ 662\ &  \ \ 3.260\ 483\ 758 & \ \ 3.553\ 390\ 863 &\ \ 4.764\ 908\ 410 \\
$(1\times 1):( 4\times 4)$ & \ \ 2.451\ 083\ 242\ & \ \ 2.928\ 442\ 860\ & \ \ 3.276\ 998\ 285 & \ \ 3.562\ 314\ 883  &\ \ 4.739\ 553\ 252 \\
\hline

$(2\times 2):( 2\times 2)$&\ \ 2.505\ 450\ 909\ & \ \  3.024\ 382\ 957\ &  \ \  3.481\ 055\ 307 & \ \ 3.717\ 691\ 692&\ \ 5.016\ 332\ 520\\
$(2\times 2):( 3\times 3)$& \ \ 2.570\ 143\ 984\ & \ \ 3.082\ 166\ 484\ & \ \ 3.454\ 087\ 416 &\ \ 3.757\ 519\ 846&\ \ 5.004\ 155\ 712  \\
$(2\times 2):( 4\times 4)$ & \ \ 2.595\ 404\ 635\ & \ \ 3.098\ 624\ 716\ &\ \ 3.378\ 293\ 046 &\ \ 3.761\ 399\ 505&\ \ 4.984\ 524\ 206   \\
  \hline

$(3\times 3):( 3\times 3)$&\ \ 2.626\ 971\ 274\ &\ \ 3.133\ 002\ 727\ & \ \ 3.497\ 087\ 416 & \ \ 3.712\ 498\ 867 &\ \ 4.992\ 841\ 134\\
$(3\times 3):( 4\times 4)$&\ \ 2.648\ 818\ 511\ & \ \ 3.147\ 204\ 863\ & \ \ 3.416\ 364\ 328 & \ \ 3.796\ 037\ 357 &\ \ 4.973\ 931\ 010 \\
  \hline

$(4\times 4):( 4\times 4)$&\ \ 2.669\ 262\ 336\ & \ \ 3.160\ 721\ 132\ &\ \ 3.598\ 289\ 910  &\ \ 3.639\ 241\ 821 &\ \ 4.954\ 642\ 401  \\
     \hline
\end{tabular}
\label{asymmkagomepotts}
\end{table}
 
 Critical thresholds  for the Potts model on kagome-type subnet lattices  computed from 
(\ref{conjecture}) are tabulated in Table \ref{asymmkagomepotts}.
For  the kagome 
lattice itself, for example,
we have  $A=A'=1,  B=B'=v, C=C'=3v^2+v^3, v=e^K-1$, and (\ref{conjecture}) gives  the critical frontier  
\be
v^6+6v^5+9v^4 -2qv^3-12qv^2 -6q^2v -q^3 =0, \quad \quad{\rm (kagome \> lattice)}.\label{pottskagome}
\ee 
The critical frontier (\ref{pottskagome}) for the kagome lattice was first obtained 
by the present author  some 30 years ago \cite{wu79,remark} by using  the
homogeneity assumption described in Sec. \ref{IIIG}.   
 Comparison of the thresholds computed from (\ref{pottskagome}) for $q=1,3,4$
 with Monte Carlo renormalization group findings has shown that the accuracy of (\ref{conjecture})
is    within one part in $10^5$ \cite{wuhu}.
  
\subsection{Site and site-bond percolation}
We now apply (\ref{conjecture}) to site as well as mixed site-bond percolation.
 First we show that (\ref{conjecture}) is exact in some instances.

{\it 1. Site percolation on the 3-12 and kagome lattices}.

\begin{figure}[htpb]
\includegraphics[scale=0.3]{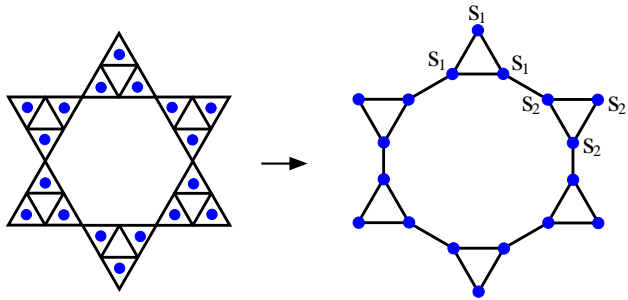}
\caption{Site percolation on the 3-12 lattice.}
\label{312siteper}
\end{figure}

 The 3-12 lattice is the lattice shown in the right panel of Fig. \ref{312siteper}. 
To formulate  site percolation on the
3-12 lattice, we consider
 the reference \ $(2\times 2):(2\times 2)$ \  lattice  with
pure 3-site interactions shown in the left.  
Let the 3-site interactions of the up- and down-triangles be, respectively,  $M_1$ and $M_2$. One finds
\bea
A&=&q^3+3\,qm_1, \quad B=m_1^2, \quad C=m_1^3, \nonumber \\
A'&=&q^3+3\,qm_2, \quad B'=m_2^2, \quad C'=m_2^3,
\eea
where $m_1=e^{M_1}-1$, $m_2=e^{M_2}-1$.  Setting $q=1$,  $m_1=s_1/(1-s_1)$, $m_2=s_2/(1-s_2)$, where $s_1$
and $s_2$ are the respective site
occupation probabilities   for the
3-12 lattice, the critical frontier (\ref{conjecture}) gives
\be
1-3(s_1s_2)^2+(s_1s_2)^3=0,\quad\quad (3-12{\rm \>\>site\>\> percolation}).
\ee

 For $s_1=s_2=s$, this yields the known   \cite{sykesessam,sudingziff99} critical frontier $1-3s^4+s^6=0$, or
 \be
s_c^{3-12}  = \sqrt {1-2\sin(\pi/18)}= 0.807\ 900\ 076. \label{312criticalpoint}
\ee
Using the relation $ s_c^{kag}= \big(s_c^{3-12}\big)^2 $ \cite{sudingziff99},
  we have therefore derived  the exact kagome and 3-12 site percolation thresholds, 
and demonstrated that
(\ref{conjecture}) is exact in this instance.
 
{\it 2. Site percolation on the martini lattice}.
\begin{figure}[htpb]
\includegraphics[scale=0.38]{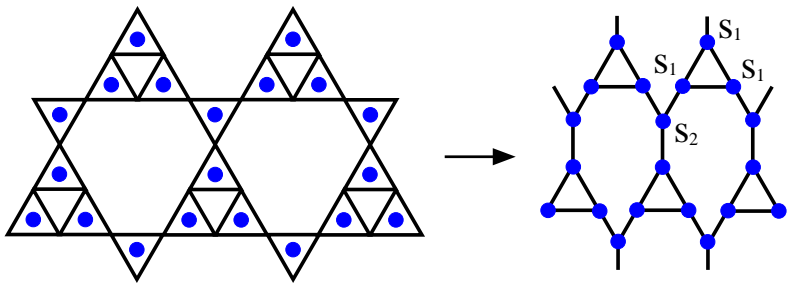}
 \caption{Site percolation on the martini lattice.}
\label{martini}
\end{figure}

The martini lattice \cite{scullard06}
is the lattice shown in the right panel
of Fig. \ref{martini}. To generate a site percolation on the martini lattice,
we start from the \ $(1\times 1):( 2\times 2)$ \ reference lattice with 3-site interactions
 shown in the left. 
 Denote the 3-site interactions of up- and down-pointing
triangular faces by $M_1$ and $M_2$ respectively and write $m_1=e^{M_1}-1$, $m_2=e^{M_2}-1$. We  have
  \bea
A&=& q^3+3\, qm_1, \quad B=m_1^2, \quad C= m_1^3 \nonumber\\
A'&=& 1, \hskip2cm \ B'=0, \quad \ \ C'=m_2.
\eea
Setting $q=1$, $m_1=s_1/(1-s_1),\,m_2=s_2/(1-s_2),$ with $s_1$ and $s_2$ the respective
site occupation probabilities,  (\ref{conjecture}) gives  the 
critical frontier
\be
1-(3\,s_1^2-s_1^3)s_2=0, \quad\quad ({\rm martini\>\>site\>\>percolation}). \label{s}
\ee
This is a   known  exact result \cite{scullardziff,kondor,ziff06}, and is 
 another example that the critical frontier (\ref{conjecture}) is exact.
For $s_2=1$, the percolation reduces to that on the kagome lattice, 
and (\ref{s}) gives  the threshold (\ref{kagsitecriticalpoint}). 
For uniform occupation probability
$s_1=s_2=s$, (\ref{s}) becomes $1-3s^3+s^4=0$ and gives the exact
solution $s_c^{martini} = 0.764\ 826\ 486$.

{\it 3. Site-bond percolation on the honeycomb lattice}. 

No exact result is known for the site and site-bond percolation on the honeycomb lattice.
Owing to the intrinsic interest of a percolation process on  a simple Bravais lattice,
the problem of honeycomb site percolation 
has attracted considerable attention for many years. There now exists a host of highly precise
numerical estimates on the threshold for site percolation   on the honeycomb lattice  \cite{sudingziff99,ziffgu09,fengdengbloete08}. 
\begin{figure}[htpb]
\includegraphics[scale=0.3]{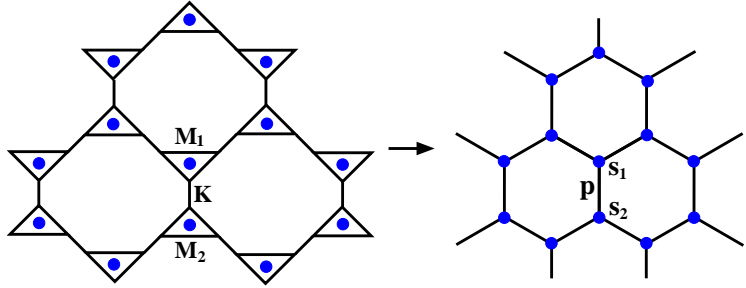}
 \caption{Site-bond percolation on the honeycomb lattice.}
\label{honeycomb}
\end{figure}

Consider the more general mixed site-bond percolation on the honeycomb lattice with site occupation probabilities 
$s_1$ and $s_2$  and 
bond occupation probability $p$  shown in the right panel of Fig. \ref{honeycomb}.
The relevant reference lattice can be taken as shown  in the left 
with edge-interactions $K$ and 
3-site interactions $M_1$ and $M_2$. To make use of (\ref{conjecture}),
 we adopt the scheme of devising up- and down-pointing triangles as indicated
 in Fig. \ref{312}(b) below.  This gives 
\bea
A&=& (q+v)^3+(q+3v)m_2, \quad B=v^2m_2,\quad C=v^3m_2, \nonumber \\
A'&=&1, \hskip4.1cm B'=0, \quad \quad \> C'=m_1,
\eea
where $v=e^K-1, m_i=e^{M_i}-1, i=1,2$. Setting $q=1, v=p/(1-p), m_i = s_i/(1-s_i)$, 
we obtain from (\ref{conjecture}) the critical frontier for the mixed site-bond percolation as
\be
(3p^2 -p^3)s_1s_2 = 1, \quad\quad  ({\rm honeycomb\>\>site-bond\>\>percolation}). \label{hcsitebond}
\ee

When $s_1=s_2=1$, (\ref{hcsitebond}) is exact since it gives the known honeycomb bond percolation threshold
$1-3p^2+p^3=0$   \cite{sykesessam,essam72,essam79}. 
When $p=1$, 
  (\ref{hcsitebond}) gives the threshold
\be 
s_1s_2 =  1/ 2  \label{hcperc}
\ee
which is exact for $s_2=1$, as    
the site percolation reduces to one
on the triangle lattice with 
  the  critical point (\ref{trisiteperc}) $s_c=1/2$.  But for $s_1=s_2=s$, 
(\ref{hcperc}) gives $s_c= 1/\sqrt2 =0.707\ 106\ 781$  differing from accurate
 numerical estimates of $s_c= 0.697\ 040\ 2$ \cite{fengdengbloete08} and $s_c= 0.697\ 041\ 3$ 
\cite{ziffgu09}. The critical frontier (\ref{conjecture}) is therefore
a close approximation in this instance.

 The site-bond percolation has also been studied by
simulations by Ziff and Gu for $s_1=s_2$ \cite{ziffgu09}
and for $p=1$ \cite{ziffgu0910}. Their results indicate (\ref{hcsitebond}) works better for site occupation
probabilities $\sim 1$.

{\it 4. Site-bond percolation on the kagome lattice}.
\begin{figure}[htpb]
\includegraphics[scale=0.18]{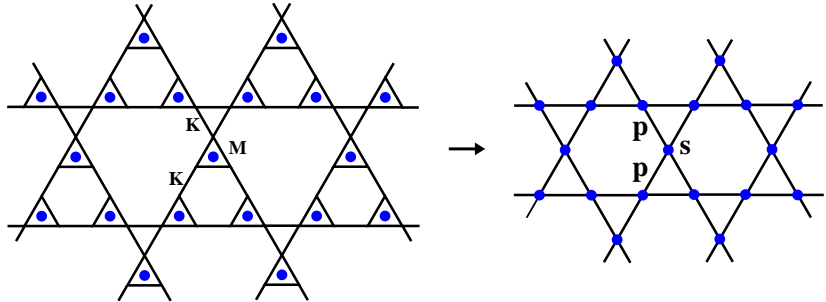}
 \caption{Site-bond percolation on the kagome lattice.}
\label{kagome-site-bond}
\end{figure}

Consider the mixed site-bond percolation on the 
kagome lattice with site and bond occupation probabilities $s$ and $p$ 
shown in the right panel of Fig. \ref{kagome-site-bond}. The
 reference lattice is shown in the left  having  edge interaction $K$ and
3-site interaction $M$. Regard the 
reference lattice as a kagome-type with the partition function
(\ref{kagomeboltzmann}). One has
 \bea
A &=& q(q^2+3m)(q+v)^3 + 3m^2(q+v)^2 +m^3, \nonumber \\
B &=& m^2 (q+v)^2 + m^3 v, \nonumber \\
C &=& m^3 (3v^2 +v^3),  \nonumber \\
A' &=& 1, \quad \quad B'=v, \quad \quad C'=3v^2+v^3, \label{kagomeSBparameter}
\eea
where $v=e^K-1, m = e^M-1$. Substituting (\ref{kagomeSBparameter}) into (\ref{conjecture}) and setting 
$q=1, v=p/(1-p), m=s/(1-s)$, one obtains the critical frontier
\bea
1+3s^2(1-3p +2p^3-p^4)  &+&s^3(-3+9p-3p^2-12p^3 +15p^4 -6p^5 +p^6) = 0, \nonumber \\
       && ({\rm kagome\>\>site-bond\>\>percolation}). \label{kagomeSB} 
\eea

For $p=1$, (\ref{kagomeSB}) becomes $1-3s^2+s^3=0$ which gives the exact critical
threshold  (\ref{kagsitecriticalpoint}) for the kagome site percolation.  For $s=1$, 
(\ref{kagomeSB}) becomes $1 - 3 p^2 - 6 p^3 + 12 p^4 - 6 p^5 + p^6 = 0$, 
or $p_c = 0.524\ 429\ 717$,  in agreement with  (\ref{kagome}}).

\subsection{The 3-12 lattice}
\begin{figure}[htpb] 
\includegraphics[scale=0.25]{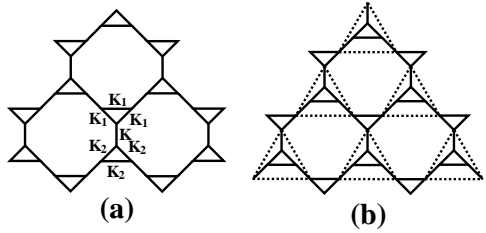}
 \caption{(a) The 3-12 Potts lattice. (b) The 3-12 lattice as an asymmetric kagome-type lattice.}
\label{312}
\end{figure}
The 3-12 lattice is the lattice shown in Fig. \ref{312}(a) with  interactions $K, K_1, K_2$.
To make use of (\ref{conjecture}) we regard the lattice as one of the 
kagome-type consisting of large up-pointing
triangles (dotted lines) and small down-pointing triangles as indicated in Fig. \ref{312}(b).
Then we have (see also Eq. (5) of \cite{wupotts06})
\bea
A&=& (q+v)^3 + 3(q+v)(q+2v) v_2
 +3(q+3v)v_2^2+(q+3v)v_2^3  \nonumber \\
 B&=& v^2[(q+v)v_2+3 v_2^2+ v_2^3]\nonumber \\
C&=& v^3(3v_2^2+v_2^3) \nonumber \\
A'&=& 1, \quad \quad B'=v_1, \quad\quad C'= 3v_1^2 + v_1^3 , \label{312abc}
\eea
where $v=e^K-1,v_1=e^{K_1}-1,v_2=e^{K_2}-1$. Substituting (\ref{312abc}) into (\ref{conjecture}), we obtain
the critical frontier (re-arranged in a symmetric form)
\be
(q+v)^3(h_1+3qv_1+q^2)(h_2+3qv_2+q^2) -3(qv^2+v^3)(h_1+qv_1)(h_2+qv_2)-(q-2)v^3h_1h_2=0,\label{312conjecture}
\ee
where $h_i = 3v_i^2+v_i^3$, $i=1,2$.
 
For the 3-12 Ising model with uniform interactions $K_I$, we set $q=2$, $ v_1=v_2=v= e^{2K_I}-1$,
and (\ref{312conjecture}) simplifies to
\be
(\sqrt 3 -1)\, v^2 -2 v -4 =0,
\ee
 yielding  the known exact critical point $e^{2K_I} = 
\frac 1 2 (3+\sqrt 3) + \sqrt {(6+5\sqrt 3)/2} = 5.073\ 446\ 135 $  in agreement
with  Utiyama \cite{utiyama} and Syozi \cite{syozi}.

For Potts model on the 3-12 lattice with uniform interaction $K$, (\ref{312conjecture}) gives
\bea
v^9+6v^8&+&3(3-q)v^7-q(32+q)v^6 -q(75+30q)v^5 \nonumber \\
 &-& q^2(111+12q)v^4-2q^3(41+q)v^3-36q^4v^2 -9q^5v-q^6=0.
\eea
This gives the critical point
\bea
e^{K_c} = v+1 &=& 3.852\ 426\ 158, \quad q=1 \nonumber\\
             &=& 5.073\ 446\ 135, \quad q=2\quad{\rm (exact\>Ising\>result)} \nonumber \\
             &=& 6.033\ 022\ 515, \quad q=3 \nonumber \\
             &=&6.857\ 394\ 828, \quad q=4. \label{312potts}
\eea
The accuracy of the prediction (\ref{312potts}) will be examined in paper II.

For bond percolation on the 3-12 lattice
we set $q=1$ and write $v=p/(1-p), v_1=p_1/(1-p_1), v_2=p_2/(p_2-p_2) $, where $p, p_1, p_2$ are the
respective bond occupation probabilities. Then
(\ref{312conjecture}) gives  the critical frontier
\bea
1-3p^2(p_1+p_1^2-p_1^3)(p_2+p_2^2-p_2^3) &+& p^3 (3p_1^2-2p_1^3)(3p_2^2-2p_2^3) =0, \nonumber \\
                     && ({\rm 3-12\>\> bond \>\>percolation}). \label{312bondperc}
\eea
This expression  has been conjectured recently by Scullard and Ziff   \cite{scullardziff}
as a non-rigorous extension of the exact bond percolation threshold of the martini lattice.
In the uniform case $p_1=p_2=p$, (\ref{312bondperc}) becomes
\be
1-p +p^2+p^3-7p^4+4p^5 =0,\quad p_c = 0.740\ 423\ 317,\label{312bond11}
\ee
which is also given by $p_c=1-e^{-K_c}$ 
using $e^{K_c}$ for $q=1$ in (\ref{312potts}). Compared to the numerical determination of
$p_c = 0.740\ 421\ 95(80)$ by
Ziff and Gu \cite{ziffgu09} and the value $p_c = 0.740\ 420\ 81$ by Parviainen \cite{parviainen07},
the accuracy of the homogeneity determination (\ref{312bond11}) 
is seen to be well within one part in $10^5$.
 
For the mixed site-bond percolation on the 3-12 lattice, it is tempting to use the kagome critical frontier
(\ref{kagomeSB}) and  replace  $s$ by $ s^2p$ as argued by Suding and Ziff \cite{sudingziff99}. 
This gives the critical frontier
\be
1+3s^4(p-3p^2+2p^4-p^5)+s^6(-3p^3+9p^4-3p^5-12p^6+15p^7-6p^8+p^9)=0.  \label{312SB}
\ee
For $p=1$ the pure site percolation, this becomes $1-3s^2+s^6=0$ which is exact. For $s=1$ the
pure bond percolation, (\ref{312SB}) gives
\be
1+3p^2-12p^3+9p^4+3p^5-15p^6+15p^7-6p^8+p^9=0 \label{312bond2}
\ee
with the solution $p_c =0.747\ 882\ 617$. The small difference between (\ref{312bond11})
and (\ref{312bond2}) reflects the approximate nature of the kagome site-bond critical frontier (\ref{kagomeSB}).

 \subsection{Critical frontier and homogeneity assumption}
\label{IIIG}
We now derive the critical conjecture (\ref{conjecture}) using a homogeneity assumption.

In the partition function (\ref{triboltzmann}),
we replace the two Boltzmann weights by 
\bea
W_\bigtriangleup (1,2,3) &=& F \times \sum_{s_1',s_2',s_3' =1}^q e^{L(\delta_{11'}+\delta_{22'}
+\delta_{33'})} e^{N\delta_{1'2'3'}}\, , \nonumber \\
W_\bigtriangledown (1,2,3) &=& F'\times \sum_{s_1' ,s_2' ,s_3' =1}^q e^{L'(\delta_{11'}+\delta_{22'}
+\delta_{33'})} e^{N'\delta_{1'2'3'}}, \label{transform}
\eea
as indicated graphically in Fig. \ref{transformation}. Equating (\ref{transform})
with (\ref{ABCdef}), we find
\bea
A&=& F\times [(q+\ell)^3+(q+3\ell)n], \nonumber \\
B&=& F\times \ell^2n\nonumber\\
C&=& F\times \ell^3n, \label{ABC}
\eea
where $\ell =e^L-1, n=e^N-1$ and similar relations for $A',B',C'$ with $\{F,\ell,n\} \to 
\{F',\ell',n'\}$,  $\ell'=e^{L'}-1,n'=e^{N'}-1$.

\begin{figure}[htpb]
\includegraphics[scale=0.25]{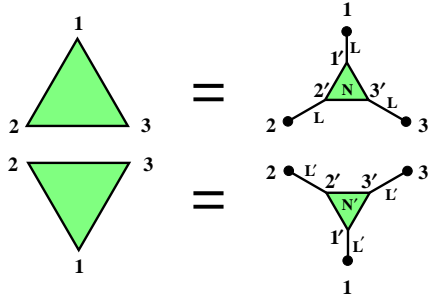}
 \caption{Graphical representation of (\ref{transform}).}
\label{transformation}
\end{figure}
\noindent
Solving  (\ref{ABC}) for $\ell, n, F$, one obtains
\bea
\ell &=& \frac C B, \nonumber \\
n &=& \frac {(qB+C)^3} {AC^2-qB^3 -3B^2C}, \nonumber \\
F &=& \frac {B^3(AC^2-qB^2-3B^2C)} {C^2(qB^2+C)^3}, \label{solv}
\eea
and similarly one obtains $\ell', n', F'$ in terms of $A', B', C'$.
The kagome-type lattice now becomes the one shown in Fig. \ref{dual}(a).

\begin{figure}[htpb]
\includegraphics[scale=0.4]{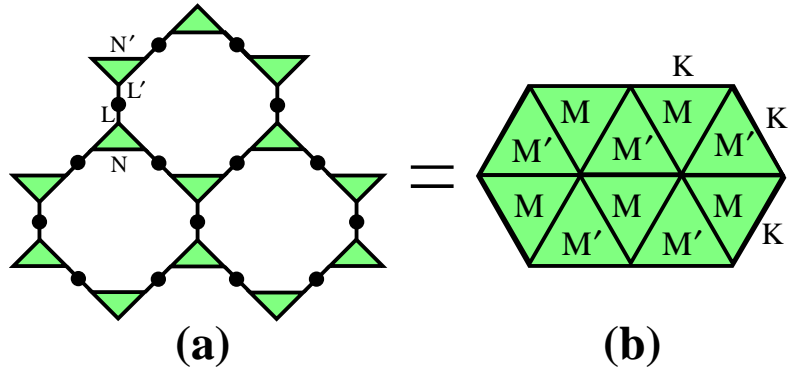}
 \caption{(a) The kagome-type lattice after transformation (\ref{transform}).
  (b) The lattice dual to (a).}
 \label{dual}
\end{figure}

The duality relation of Potts models with multi-site interactions has been formulated by Essam 
\cite{essam79} (see also \cite{wureview}). Following Essam,
the  dual to the lattice in Fig. \ref{dual}(a) is the one shown in Fig. \ref{dual}(b) with
 \bea
e^K &=&(1+q/\ell)(1+q/\ell') ,  \nonumber \\
e^M &=& 1+q^2/n, \nonumber \\
e^{M'} &=& 1+ q^2 /n'\, ,\label{dualinteraction}
\eea
where the interaction  $K$ is the dual to
the two interactions $L$ and $L'$ in series.
We therefore are led to consider the Potts model on the 
triangular lattice shown in Fig. \ref{dual}(b), where $M$ and $M'$ are 3-site interactions.
 
For $M'=0$, the partition function is $Z_{tri}(q;A,B,C)$ given by (\ref{triboltzmann})  with
\be
W_\bigtriangleup(1,2,3) = e^{K(\delta_{12}+\delta_{23}+\delta_{31})}e^{M\delta_{123}},\nonumber
\ee
or
\be
A=1,\quad B=e^K-1, \quad C= e^{3K+M}-3e^K+2\, .
\ee
The exact critical frontier in this case is known.  It is \ $qA=C$,  or
\be
e^{3K+M}-3e^K+2 = q. \label{zeroMp}
\ee
For $M'\not= 0$ the critical frontier is not known. However,  the critical frontier must be 
symmetric in $M$ and $M'$.  We now make a {\it homogeneity  assumption}  requiring $M$ and $M'$ to appear 
homogeneously in the exponent  of (\ref{zeroMp}). The simplest way to do this is
to  extend (\ref{zeroMp}) to 
\be
e^{3K+M+M'}-3e^K+2 = q\, . \label{zeroMp1}
\ee 
The substitution of expressions of $K,M$ and $M'$ in (\ref{dualinteraction}) 
and (\ref{solv}) into (\ref{zeroMp}) now leads to  (\ref{conjecture}).

\section{Summary}
We have considered the $q$-state Potts model and the related  bond, site, and mixed site-bond
 percolation
for  triangular- and kagome-type lattices. For triangular-type lattices we obtained its exact critical
frontier in the form of (\ref{tri}) without   the usual assumption of a unique transition.  We then applied 
the exact critical frontier  in various applications.
For  kagome-type lattices we obtained a new  critical frontier (\ref{conjecture}) 
by making use of a homogeneity assumption.
We established that the new critical frontier 
is exact for $q=2$ and for site percolation on the kagome,   martini, and other
lattices.  For the  Potts and bond percolation models for which there is no exact solution, 
the new critical frontier
 gives numerical values of  critical thresholds accurate to  the order of  $10^{-5}$. 
For mixed site-bond percolation, 
the homogeneity assumption gives rise to critical frontiers which are  accurate
when site occupation probabilities are $\sim 1$.

In summary, we emphasize that applications of the critical frontiers (\ref{tri}) 
and (\ref{conjecture}) are not limited to those reported
in this paper.  They can be extended to numerous other lattice 
models having a triangular or kagome symmetry, and thus they open
 the door to a host of  previously unsolved problems.

\section*{Acknowledgment}
I would like to thank Chengxiang Ding for help in the preparation of the manuscript and Wenan Guo for a 
critical reading. I am grateful to R. M. Ziff for helpful comments and suggestions,
 and for communicating on results prior to publication.

\end{document}